\documentclass[%
 reprint,
unsortedaddress,
nofootinbib,
 amsmath,amssymb,
 aps,
floatfix,
]{revtex4-2}

\usepackage{comment}
\usepackage{graphicx}% Include figure files
\usepackage{dcolumn}% Align table columns on decimal point
\usepackage{bm}% bold math
\usepackage[dvipsnames]{xcolor}
\newcommand{\specificthanks}[1]{\textit{$^{\ddag}$}}
\newcommand{\vfh}[1]
{\textcolor{blue}{#1}}

\begin{document}

\preprint{APS/123-QED}

\title{Competition Between Energy and Dynamics in Memory Formation %Multi-Stable Systems
}

%\title{Dynamical multi-stability in spring systems/ Using dynamics to guide memory formation in multi-stable systems/ Competition between energy and dynamics in memory formation}

%\author{Varda F. Hagh\textsuperscript{a,*}\thanks{vardahagh@uchicago.edu}}
\author{Varda F. Hagh\textit{$^{\ddag}$}}
\email{vardahagh@uchicago.edu}
\author{Chloe W. Lindeman\textit{$^{\ddag}$}}
\email{cwlindeman@uchicago.edu}
\author{Chi Ian Ip}
\author{Sidney R. Nagel}%
\affiliation{%
 Department of Physics and The James Franck and Enrico Fermi Institutes \\ 
 The University  of  Chicago, 
 Chicago,  IL  60637,  USA.\\
 \\
\normalsize{$^{\ddag}$ Equal Contribution}
%\normalsize{$^{\ddag}$ Equal Contribution}
\\
}%

\date{\today}% It is always \today, today,
 
%\textcolor{blue}{NOTE:  I can't figure out how to get commas in right place for Varda and Chloe in author list.  See note about combing Fig. 2 and 3. Also note changed title.  I think that Chloe went for this new title too, but we should discuss.}\\
%\textcolor{purple}{I think option 1 is nice visually, but it seems like it may be against the rules to reference the figures out of order. In that case, I think 3 or 4 are both okay. Is there a reason to limit the number of total figures? If so, maybe 4 is not so good?}}        

\begin{abstract}
Bi-stable objects that are pushed between states by an external field are often used as a simple model to study memory formation in disordered %, out-of-equilibrium 
materials. Such systems, called hysterons, are typically treated quasistatically. %even though the behavior of real multi-stable systems may depend on the timescale of forcing. 
Here, we generalize hysterons to explore the effect of dynamics in a simple spring system with tunable bistability and study how the system chooses a minimum. Changing the timescale of the forcing allows the system to transition between a situation where its fate is determined by following the local energy minimum to one where it is trapped in a shallow well determined by the path taken through configuration space. Oscillatory forcing can lead to transients lasting many cycles, a behavior not possible for a single quasistatic hysteron.
\end{abstract}

\maketitle  
\section{Introduction}

%The ability to store memories is a feature of many disordered systems.
The ability of a physical system to store information about how it was prepared --- memory --- is now recognized as being crucial for the behavior in a large variety of disordered materials~\cite{KeimPaulsen2019}. Jammed packings of soft spheres subjected to repeated cycles of shear, cyclically crumpled sheets of paper, and interacting spins in an oscillating magnetic field all form memories of how they were trained~\cite{fiocco2014encoding, fiocco2015memory, adhikari2018memory,arceri2021marginal,charbonneau2021memory, shohat2022memory}. Memory in such systems hinges on the ability to learn a pathway between metastable states of the energy landscape. It has been likened to the memory seen in a collection of bi-stable elements, called hysterons, which flip between states when an external field is raised above or below a critical value as shown in Fig.~\ref{two springs}(a)~\cite{preisach1935magnetische,keim2020global,mungan2019networks}. Although an enormous simplification from the original materials, ensembles of hysterons are able to capture some features of the memory formation seen in complex systems surprisingly well~\cite{KeimPaulsen2019, mungan2019networks, Mungan19}.  %However, as has been shown recently, the simplest hysteron models are insufficient to explain and understand complex behavior in many systems~\cite{lindeman2021multiple, keim2021multiperiodic, bense2021complex, mungan2019networks}. Here, we develop a generalized concept of hysterons that can serve as a foundation for understanding memory in cases not amenable to traditional approaches, particularly in systems with inherent dynamics or out of mechanical equilibrium. 
%Although not apparent ``by eye'', these memories can be read out using a specific protocol. Generally speaking, a cusp in the readout curve indicates the amplitude at which the system was trained and therefore allows a direct measurement of how the system was prepared.

However, such hysterons fail to capture certain features of real systems~\cite{mungan2019networks, lindeman2021multiple, keim2021multiperiodic, bense2021complex}. As long as the hysterons are independent, for example, %it takes only one cycle to train the system;
the configuration produced at the end of the first cycle is guaranteed to be the same as that found after subsequent cycles of the same amplitude (since each hysteron separately has this property).  By contrast, cyclically sheared packings can take many cycles to train, and can even exhibit a multi-period response~\cite{lavrentovich2017period} in which the periodicity of the response is an integer multiple of the driving period as first demonstrated in systems with friction~\cite{Royer15}. Recent work has shown that generalizing the simple idea of a hysteron as an independent two-state object by adding interactions can result in long training times and multi-period responses~\cite{lindeman2021multiple, keim2021multiperiodic}.

Here, we generalize the behavior of hysterons in a different way: by studying the effect of dynamics. Starting from a two-spring configuration that gives rise to a symmetric double-well potential, %(section~\ref{Two spring section}), 
we add features one at a time to uncover the criteria for landing in one basin or the other. %As a result of this competition, the qualitative behavior changes just by changing the timescale of forcing. 
%Although the system is simple enough to understand analytically in some cases, the resulting dynamics are rich with complexity and provide insight into how introducing new timescales might affect the behavior of hysteron-like objects and the complex, disordered systems they aim to describe.
%In section~\ref{constant f_w section} 
When a symmetry-breaking third spring is added, the behavior is determined by a competition between the timescale of applied forcing %changing the shape of the potential 
and the timescale of inherent system dynamics that relaxes the system to lower energy. There is a crossover, depending on forcing velocity, between \textit{energy-dominated} and \textit{path-dominated} selection criteria.
%in section~\ref{cyclic forcing section} we 
Following from this, for oscillatory driving we find a critical frequency which separates the two regimes. Finally, we characterize the effect of allowing the system to age by slowly evolving the spring stiffnesses. %(section~\ref{aging section}).

%In a sense, 
%The rugged energy landscapes characteristic of these complex, disordered systems make this memory formation possible since the effect of training is to push the system from its initial configuration into a new energy minimum. At the same time, the complexity of these energy landscapes can obscure the mechanism of memory formation: if the system were to fall into a new minimum with every small push, we might expect it to immediately forget anything it learns.

%The rugged energy landscapes characteristic of complex, disordered systems suggest why hysterons might be useful in modelling memory formation. Taking a low-dimensional cut through two neighboring wells in parameter space will generically give rise to a double-well energy potential that is morphed by an external field (for example, shear). 

%Double wells morphed by an external field so that they display hysteretic behavior, as shown in Fig.~\ref{hysteron}, have been used to describe a variety of disordered systems. 
%with the system remaining in one well as the strain $\gamma$ increases until $\gamma^+$ and not returning to the original well as the strain decreases until $\gamma^- < \gamma^+$. 

%\section{Results}

\section{Unbuckled-to-buckled transition for two springs}
\label{Two spring section}

\begin{figure}[h!]
\centering
\includegraphics[width=8.6cm]{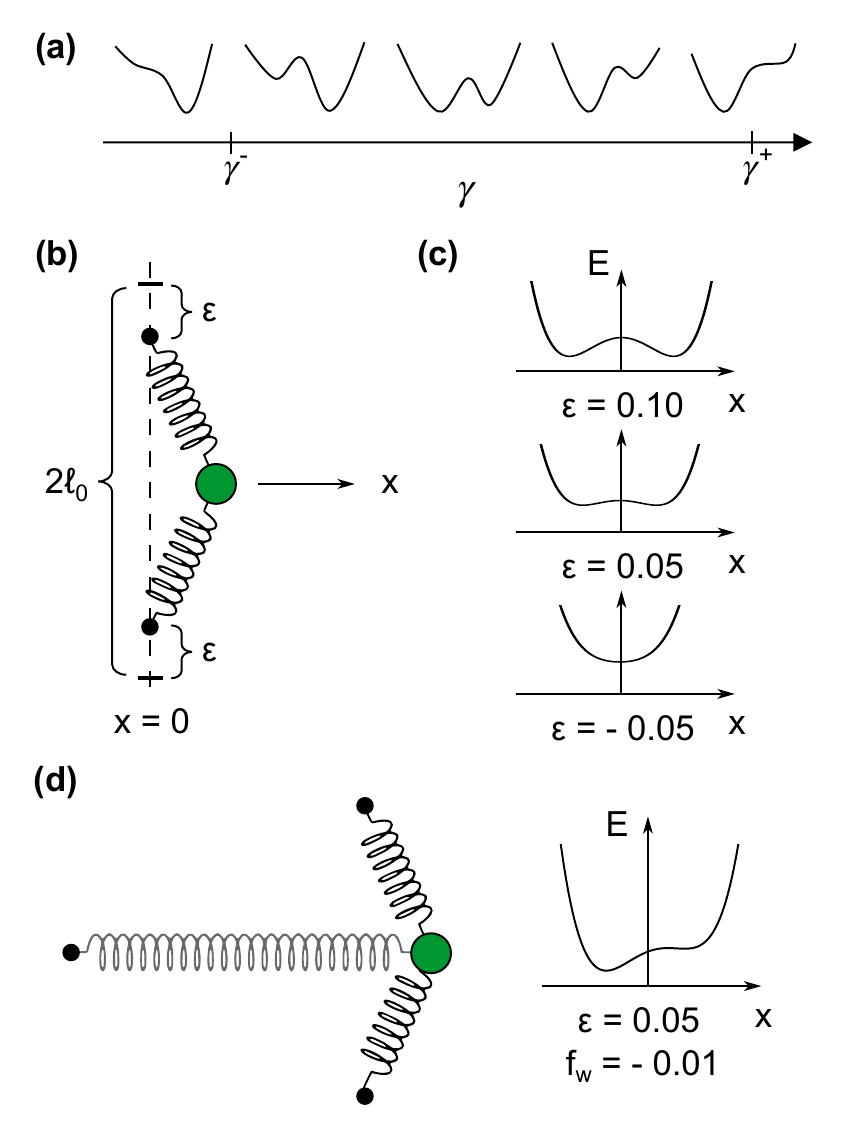}
\caption{
(a) Double-well model of a hysteron. Under an applied strain, $\gamma$, the landscape changes: one well disappears at low strain and the other disappears at high strain.
(b) Schematic of the two-spring system. Outer nodes are shown as black dots. Middle node is shown as a green circle. 
%(c) Two-spring system with the springs (dashed lines) fixed at one end (black dots) and connected to a node at the other end (green circle). Shading shows the associated energy for all possible positions of the middle node. Orange dots mark the location of the two minima. 
(c) Energy versus position of the middle node. For large $\epsilon$, the two wells are deep. For small $\epsilon$, the wells become smaller and the minima are closer together; they coalesce when $\epsilon \rightarrow 0$. For $\epsilon \le 0$, there is only one minimum.  
(d) Three-spring system with $f_w \ne 0$ showing energy versus middle-node position.  In the energy diagram, the force is pulling to the left (in the negative $x$-direction).
}
\label{two springs}
\end{figure}

In two dimensions, two identical harmonic springs with rest lengths $\ell_0$ and stiffnesses $k$ can be connected by a single node to produce a bistable system as shown in Fig.~\ref{two springs}(b,c). The central-node location is the only variable since the positions of the two outer nodes are explicitly controlled. We restrict the motion of those outer nodes to be symmetric about the $x$-axis so that the middle node moves only in one dimension along $x$. The distance of a given outer node from its position when the two outer nodes are exactly $2 \ell_0$ apart is given by $\epsilon$.
%The extent to which each spring is compressed, $\epsilon$ \vfh{[$\epsilon$ is not the amount by which the spring is compressed. $\ell^2 = (\ell_0 - \epsilon)^2 + x_0^2$. So $delta \ \ell = \ell - \ell_0$ which is how much the spring compressed does not only depend on $\epsilon$]}, is the distance of a given outer node from its position when the two outer nodes are exactly $2 \ell_0$ apart. 
The energy is the sum of the spring energies: 
$$E = k(\sqrt{x^2 + (\ell_0-\epsilon)^2}-\ell_0)^2$$ 
which to first order in $\epsilon$ and fourth order in $x$ is: 
\begin{equation}
E \approx k \Big( \frac{1}{4 \ell_0^2}x^4 - \frac{\epsilon}{\ell_0}x^2 \Big).
\label{two-spring energy}
\end{equation}
The energy is symmetric %$x \rightarrow -x$ 
around $x=0$, the center of symmetry of the energy landscape. When $\epsilon > 0$, the quartic and quadratic terms are of opposite sign and the energy is given by a bistable (double-well) potential.  This corresponds to a buckled configuration. %hence the bistable nature of the system. 
For $\epsilon < 0$, the springs are stretched and there is only a single minimum.   %is along that axis and we need only worry about one-dimensional motion. 

We study the behavior of this system with over-damped dynamics. The $x$-velocity of the middle node is given by $v_{m} = f/\beta$ where $f$ is the $x$-component of the total force from all springs and $\beta$ is a damping coefficient. In addition to the spring forces, an $x$-velocity of the outer nodes, $v_{o,x}$, will also influence the motion of the middle node. %Noting that over-damped dynamics introduces a preferred frame of reference (determined by the medium producing damping), we
We choose a reference frame in which the outer nodes are stationary in the $x$-direction. %and the medium moves in the opposite direction. 
By its definition the position $x=0$ then also remains stationary so that the middle node moves with respect to $x=0$ with an additional velocity: $\Delta v_{m} = -v_{o,x}$.  This movement, due solely to this motion of the origin, is \textit{in addition} to the velocity caused by the springs themselves. With over-damped dynamics, this can be treated as if there were an additional effective force in the $x$-direction $f_{\textrm{eff}}=-\beta v_{o,x}$:
\begin{equation}
E_{v_x} \approx k \Big( \frac{1}{4 \ell_0^2}x^4 - \frac{\epsilon}{\ell_0}x^2 \Big) + \beta v_{o,x} x.
\label{moving energy}
\end{equation}

We probe the transition from one to two minima by bringing the two outer nodes together in the $y$-direction at a fixed velocity $v_y$.  %In general, we expect such a symmetry to mean that, as 
Starting from the stretched state with one minimum, $\epsilon$ changes from negative to positive as the two nodes approach one another, causing the initial minimum to separate into two distinct minima. If the motion is only along the $y$-direction (co-linear motion of the two outer nodes), the system chooses a minimum randomly (if there is any noise) or remains stuck in the unstable equilibrium position $x=0$. 

However, %this is not necessarily the case 
if the outer nodes %break this symmetry %are not restricted to move in a co-linear fashion but may by 
move in the $x$- as well as the $y$-direction, %any motion in the $x$-direction 
%Unless otherwise noted, we will stick to this convention so that $x=0$ is always defined by the line between the two outer nodes. 
the path of the outer nodes dictates into which minimum the system will eventually come to rest.  One can see already that this two-state system is different from the conventional hysteron; the \textit{path} of applied boundary motion, not just the resulting \textit{energy landscape}, determines the configuration of the system.

%Because the minimum is chosen by the path of the boundaries even while the shape of the energy landscape remains symmetric, we call the system's choice of minimum ``path dependent."

\section{Addition of a weak third spring}
\label{constant f_w section}

One can break the symmetry of the two-spring system by %adding an extra force acting in one direction, for example by 
attaching a third, weak spring to the middle node so that it applies an additional force along the $x$-axis, as shown in Fig.~\ref{two springs}(d). If the other end of the third spring is pinned so that the equilibrium position is far away, the force will be approximately independent of position, and the modified energy will be given by 
\begin{equation}
E \approx k \Big( \frac{1}{4 \ell_0^2}x^4 - \frac{\epsilon}{\ell_0}x^2 \Big) - f_w x,
\label{three-spring energy}
\end{equation}
where $f_w$ is the small force due to the weak spring. 
The form of this equation is identical to that of Eq.~\ref{moving energy}, so we can include any boundary-node motion in the $x$-direction as an effective weak spring force: $f_{\textrm{eff}} = f_w - \beta v_{o,x}$. %, with $v_{x,\textrm{eff}} = 0$. 
%In what follows, we will fix the boundary node positions and show the effect of changing the weak force.

% This paragraph no longer really makes sense; we don't use the results here.
% I moved it to the detailed analysis section below.
\begin{comment}
We can calculate for this asymmetric case with fixed $\epsilon$ ($i.e.$, $v_y = 0$) the value of $f_w$ at which a second minimum just forms. We do this by setting the maximum force of the two-spring system equal and opposite to the force provided by the weak spring at $x_c$, the critical position where the force from the strong springs is a maximum. We find 
%$$ -\frac{d^2E}{dx^2} = k \Big( \frac{3}{\ell_0^2}x^2 - \frac{2\epsilon}{\ell_0} \Big) = 0 $$
$$ x_c = \pm \sqrt{\frac{2}{3} \ell_0 \epsilon}$$
and
\begin{equation}
f_c \sim k \Big( \frac{\epsilon^3}{\ell_0} \Big)^{1/2}.
\label{critical epsilon}
\end{equation}
\end{comment}

%One interpretation of the above equation is that t
This simple model of a hysteron has two distinct mechanisms that can compete to determine the effective force. If the effective force is dominated by the weak spring, then the behavior will be ``energy dominated,'' with the landscape changing slowly enough that the energetics determine which well is chosen. If, on the other hand, the velocity of the boundary nodes dominates, then the behavior will be ``path dominated'': the energy landscape will change too quickly for the system to keep up with the local minimum and so it will become trapped in a state determined by the path of the boundaries. 

We can make the crossover between energy-dominated and path-dominated outcomes explicit by finding the critical $x$-velocity of the boundary nodes that leads to zero effective force: $f_{\textrm{eff}} = f_w - \beta v_{o,x} = 0$. %, or $v_x = f_w/\beta$. 
If we start with $\epsilon < 0$ at $x = 0$ and bring the nodes together at a constant velocity, we find that for $v_{o,x} < f_w/\beta$, the system is energy-dominated and ends up in the global minimum; for $v_{o,x} > f_w/\beta$, the system is path-dominated and becomes trapped in the shallower minimum.

%From this analysis, we determine that the crossover between when path dependence and energy dependence for choosing the configuration of the system occurs when $f_{\textrm{eff}} = f_w + \beta v_x = 0$ or $v_x = -f_w/\beta$.  That is, this simple model of a hysteron has two mechanisms by which it can choose its local minimum depending on the velocity of forcing.  If we start with $\epsilon < 0$ at the equilibrium position, \textcolor{blue}{Check that this is sufficient.}for $v_x < -f_w/\beta$, the system ends up in the lowest energy configuration while for $v_x > -f_w/\beta$, the eventual configuration will be determined by the velocity of the outer nodes.

\section{Oscillatory forcing}
\label{cyclic forcing section}

Now that we have understood the basics of these three-spring systems, we can bring them closer to the hysteron picture by modulating the weak force with time so that the bi-stable system is subjected to oscillatory forcing: %We introduce an oscillating external field by allowing the weak force due to the third spring to oscillate in time: 
\begin{equation}
    \label{eq:f_w}
    f_w(t) = f_0 + A \sin(\omega t).
\end{equation}
Recall that independent quasi-static hysterons are completely trained in one cycle. %: the state of each hysteron after the first cycle is identical to its state after subsequent cycles of the same amplitude~\cite{REF given in introdction}. 
However, in this realization of a hysteron with dynamics, it can take many cycles before the system reaches its steady-state behavior. %as viewed stroboscopically at the end of each cycle.
\begin{comment}

\begin{figure}%[h!]
\centering
\includegraphics[trim= 0cm 0cm 0cm 0cm, width=6.5cm]{omega_c.pdf}%
\caption{a) Trajectories (positions vs time) of three systems with fixed driving amplitude, $A=10^{-4}$, but different values of the driving frequencies $\omega=0.08$ (red), $0.095$ (blue), and $0.11$ (black). The red and blue curves take multiple cycles before falling to the global minimum; the black curve does not fall to the global minimum in the time shown. The critical frequency for this system is $\omega_c=0.103$. b) The critical frequency, $\omega_c$, at which the crossover time diverges, for a variety of different oscillation amplitudes, $A$. The slope of dashed fitted line is $1.0020 \pm 0.0005$, showing that the scaling of $\omega_c$ with $A$ matches the analytical results given by Eq.~\ref{eq:omega_c}. The error bars are smaller than data markers.}
\label{fig:omega_c}
\end{figure}   
\end{comment}

We analyze the case where there are two stable minima at time $t=0$ and the middle node begins in the energetically unfavorable well. If both wells remain stable at all times, the middle node will never leave its initial well. However, if for some fraction of the cycle, the higher-energy minimum disappears, the middle node may escape to the deeper minimum. 

\begin{figure}[h!]
\centering
\includegraphics[ width=6.0cm]{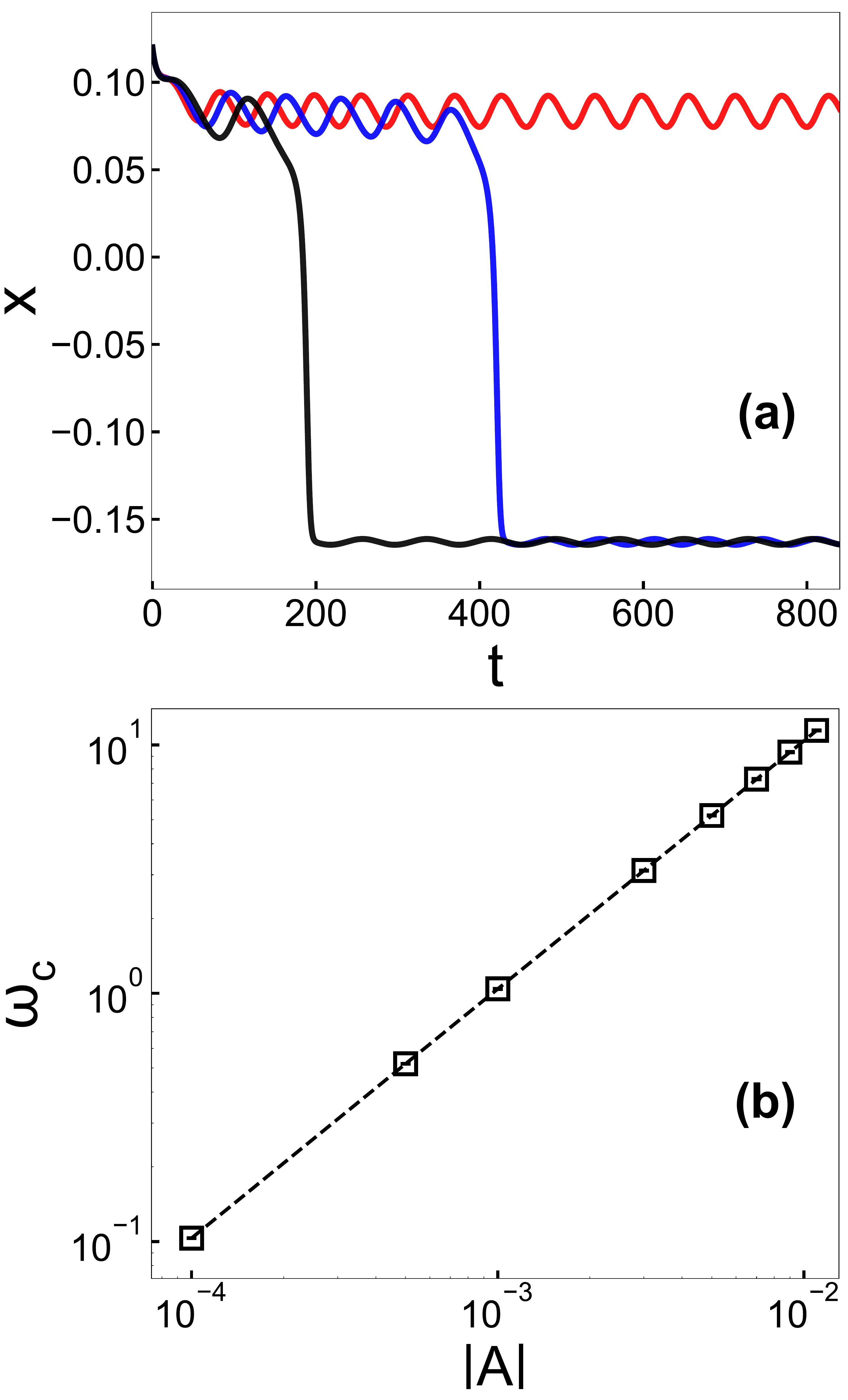}%
\caption{(a) Position versus time for three identical systems driven at different frequencies. %fixed driving amplitude, $A$ and weak force $f_0$ \vfh{[weak force is not equal to $f_0$.]}, but different driving frequencies.
For the values of $f_0$ and $A$ used, the black $\omega=0.08$ and blue $\omega=0.095$ curves take multiple cycles before falling into the global minimum; the red $\omega = 0.11$ curve is above the critical frequency at which the crossover time diverges and remains in the upper well. For this system, $\omega_c \approx 0.103$. (b) The measured critical frequency $\omega_c$ versus the magnitudes of oscillation amplitudes $|A|$. The slope of the dashed fitted line is $\alpha = 1.001 \pm 0.004$. %showing that the scaling of $\omega_c$ with $A$ matches the analytical results given by Eq.~\ref{eq:omega_c}.
The error bars are smaller than data markers.}
\label{fig:omega_c}
\end{figure}

We simulate this behavior by calculating the net force on (and hence velocity of) the middle node and updating its position accordingly. Length and time are measured in units of $\ell_0$ and $\beta / k$ respectively.
%by iterating the equation of motion using \textcolor{blue}{units where the units of length and time are meters and seconds, respectively.}
%The velocity, proportional to the net force, is computed and used to find the node's new position at each time step. %This process is repeated $1000$ times.  
The node position versus time is shown in Fig.~\ref{fig:omega_c}a. 
At low frequency $\omega$, the system initially remains in the metastable minimum but after several cycles escapes to the global minimum. %We can imagine measuring two different critical frequencies: $\omega_0$, the frequency that tells us whether a system will fall into the stable well after a single cycle, and
As $\omega$ increases, it takes longer for the node to reach that global minimum. This time diverges at a critical frequency $\omega_c$, above which the system never falls into the deeper well. %As shown in Fig.~\ref{fig:omega_c}b, $\omega_c$ depends on the amplitude $A$.  

We find the crossover time at which the middle node falls into the lower energy well.  The critical frequency, $\omega_c$, is determined by the limit when the crossover time versus $A$ diverges. Unless otherwise noted, we have chosen $\beta = 0.10$, $\epsilon = 0.01$, $A=-10^{-4}$ and $f_0=-0.00108$, and have set the initial middle-node position to be $x = 0.12$. %, near to the minimum of the higher-energy well. %For $\omega$ near $\omega_c$, the crossover time $t_c$ becomes large and we can fit for $\omega_c$ by finding the best value for which $log(1/T_c)$ becomes linear against $log(\omega_c - \omega)$. 
As shown in Fig.~\ref{fig:omega_c}b, $\omega \propto A^{\alpha}$
with $\alpha = 1.001 \pm 0.004$.

To calculate the scaling of $\omega_c$ with $A$, 
%under a set of simplifying assumptions. %In practice, however, $\omega_0$ is not an easily defined value since it requires setting a threshold position that separates the two wells and depends on the initial position of the middle node. We will therefore restrict our analysis to determining the ultimate fate ($t \rightarrow \infty$) of the middle node rather than calculate the exact time it crosses to the global minimum.
% the asymptoti will
we need to find the frequency at which the motion in one half of the cycle is exactly undone by the motion in the other half. Any longer of a period and the system will start to creep progressively toward the deeper well; any shorter and the system will be stuck eternally. %This situation therefore defines the critical frequency $\omega_c$. 

\begin{figure}%[h!]
\centering
\includegraphics[width=8.6cm]{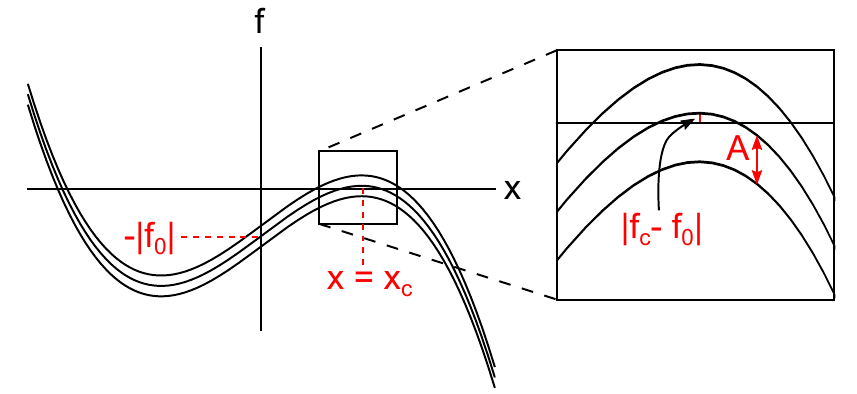}
\caption{
Net force $f$ on the middle node versus its position for $f_w = f_0 - A$, $f_w = f_0$, and $f_w = f_0 + A$ with $0 < |f_c - f_0| < A < |f_c|$. Note in this case that $f_0$ and $f_c$ are both negative; $|f_c|$ is just barely larger than $|f_0|$. A zero of the force corresponds to a local minimum or maximum of the energy; inset shows the location of the energy minimum which is stable only during part of each cycle. 
}
\label{fig:force}
\end{figure}

We can calculate $\omega_c$ analytically under a set of simplifying assumptions. (i) We replace the sine function in Eq.~\ref{eq:f_w} with a square wave so that $f_w$ takes on only two discrete values $f_0 + A$ and $f_0 - A$, shown schematically in Fig.~\ref{fig:force}. %The behavior of the response %, both quantitative and qualitative, will depend on the frequency of oscillation. Under this simplification, we can calculate
%In order to calculate $\omega_c$ analytically, we %will make explicit our criteria for oscillation. As stated above, we take 
(ii) We define a force $f_c$ to be the value at which a second minimum just forms and assume $0 < f_c - f_0 \ll f_c$ so that the time-averaged potential is just barely stable and $f_c - f_0 \ll A$ so that the potential is relatively flat ($i.e.$, variation in the force is small compared to its magnitude in the vicinity of the critical position $x_c$ where the force from the strong springs is a maximum). (iii) We assume small oscillation amplitudes: $A \ll f_c$. Fig.~\ref{fig:force} provides a graphical interpretation of these statements. 

%With this final assumption of flatness, we can approximate the displacement of the middle node as the time spent in half of the cycle $T/2$ times a characteristic velocity of the middle node $v \sim \langle F \rangle/\beta$, where $F$ is the force averaged over position. (Note that the exact displacement is given by the force averaged over time, which is much more difficult to compute. However for low spatial variation in F they approach one another.) Finally, we Taylor expand the force as a quadratic around $x_c$:

With the assumption of flatness in (ii), we can approximate the displacement of the middle node as the time spent in half of the cycle $T/2$ times a characteristic velocity of the middle node $v \sim \langle f \rangle/\beta$, where $\langle f \rangle$ is the force averaged over time. The time average can be well approximated as a position average for low spatial variation in $f$, which is the case we are interested in here. Finally, we Taylor expand the force as a quadratic around $x_c$:

\begin{equation}
f(x) \approx (f_0 \pm A) - f_c - \frac{3 k}{\ell_0^2}(x-x_c)^2 x_c.
\label{taylor approx}
\end{equation}

We estimate the critical frequency by determining the size of a limit cycle centered at $x_c$ and then finding the period of a cycle with that size. For such a cycle, 
\begin{multline*}
    \int_{x_c - a}^{x_c+a}dx \Big[  (f_0 + A) - f_c  - \frac{3 k}{\ell_0^2}(x-x_c)^2 x_c \Big] \\
    = \int_{x_c + a}^{x_c-a}dx \Big[ (f_0 - A) - f_c  - \frac{3 k}{\ell_0^2}(x-x_c)^2 x_c \Big]
\end{multline*}
\begin{equation*}
    \Rightarrow \int_{x_c - a}^{x_c+a}dx \Big[ f_0 - f_c - \frac{3 k}{\ell_0^2}(x-x_c)^2 x_c \Big] = 0,
\end{equation*}
where $2a$ is the size of the cycle. 
%We have thus simplified our calculation to an integral over the time-averaged force curve; as a result, the amplitude of oscillation $A$ does not play a role in this part of the calculation. We will see that it appears when we convert this limit cycle size to the associated period and hence frequency.
This can be solved for $a$ with the quadratic form of the force given in Eq.~\ref{taylor approx}. Substituting in $x_c = \sqrt{2 \epsilon \ell_0 /3 }$, which can be found by setting the derivative of the force equal to zero, gives 
\begin{equation*}
    a \sim \sqrt{\frac{f_0 - f_c}{k}} \Big( \frac {\ell_0^3}{\epsilon} \Big)^{1/4}.
\end{equation*}

To get a cycle of this size, we need the middle node to travel distance $a$ in a quarter cycle: $a = f T/4 \beta$, where $T$ is the period and $f$ is the characteristic force and hence determines the characteristic velocity $f/\beta$. By our assumption of flatness, the force near $x_c$ at any point in the cycle is very close to $\pm A$, so we find 
\begin{equation*}
    T_c \sim \frac{\beta a}{f}
\end{equation*}
%\begin{equation*}
%    \Rightarrow T_c \sim  \frac{\beta}{f}\sqrt{\frac{f_0 - f_c}{k}} \Big( \frac {\ell_0^3}{\epsilon} \Big)^{1/4}
%\end{equation*}
\begin{equation}
    \label{eq:omega_c}
    %\Rightarrow 
    \omega_c = \frac{2\pi}{T_c} \sim \frac{|A|}{\beta} \sqrt{\frac{k}{f_0 - f_c}} \Big( \frac {\epsilon}{\ell_0^3} \Big)^{1/4}.
\end{equation}
This is consistent with the exponent $\alpha$ obtained in the simulations. In fact, we see from Fig.~\ref{fig:omega_c}b that $\alpha \sim 1$ even when $A$ is extended to values larger than $f_0$, well outside the regime of the calculated exponent.

\section{Directed Aging}
\label{aging section}
Existence of a critical frequency indicates that the behavior of a dynamic hysteron is determined by a competition between the timescale of external driving, set by $\omega$, and the time it takes to travel a given distance, set by $\beta$ and the net force from the springs at any given moment. Any process that alters one of these timescales can therefore change the behavior of the system. 

One example of such a process is directed aging in which local properties of a spring network, such as the spring constants or bond lengths, evolve in response to the stresses imposed on each bond~\cite{pashine2019directed, hexner2020, hexner_pashine2020}. This leads to changes in the global elastic response of the system. 

In the dynamic hysterons described above, each spring, $i$, undergoes a strain $\delta \ell_i $ when the system is driven. We evolve the spring constants at an aging rate  $q$ according to the energy stored in each bond: 
\begin{equation}
\label{eq:directed_aging}
k_i (t+1) = k_i(t) -  q\ k_i (t)\ \delta \ell_{i}^{2}.
\end{equation}

\begin{figure}[h!]
\centering
\includegraphics[trim= 0cm 0cm 0cm 0cm, width=6.0cm]{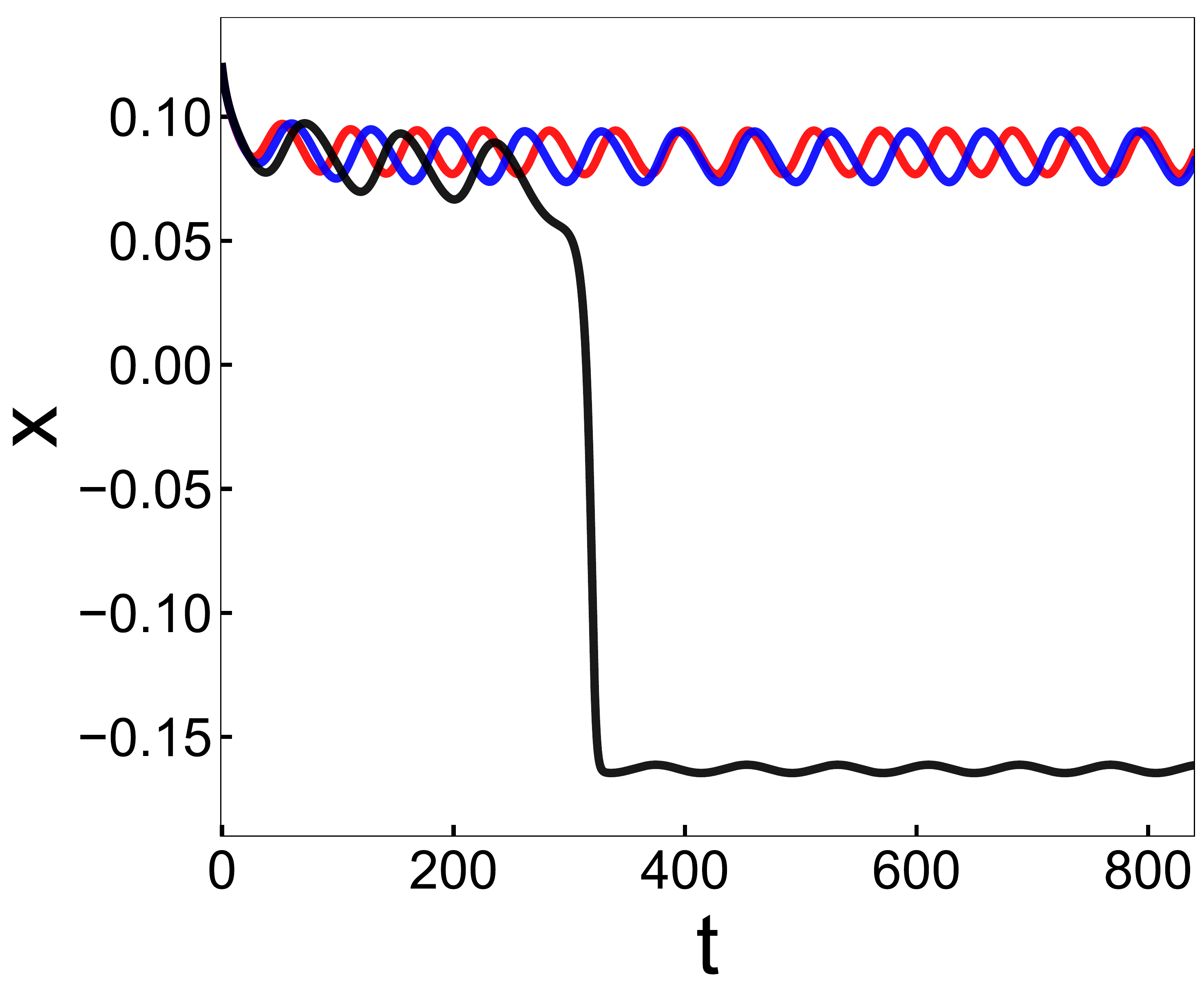}%
\caption{Trajectories of the three systems shown in Fig.~\ref{fig:omega_c}a %with similar fixed driving amplitude, $A=10^{-4}$, and values of the driving frequencies $\omega=0.08$ (red), $0.095$ (blue), and $0.11$ (black) 
after the system has been aged. Now only the black curve falls into the global minimum because the critical frequency has decreased to $\omega_c \approx 0.088$.}
\label{fig:omega_c_aged}
\end{figure}

%The timescales that depend on the stiffness change dramatically. 
Fig.~\ref{fig:omega_c_aged} shows the trajectories for the same system at the same frequencies depicted in Fig.~\ref{fig:omega_c}a after the system is aged. 
Starting with $k = 1$ in the strong springs and $k_w = 10^{-3}$ in the weak spring, we update the stiffness values according to Eq.~(\ref{eq:directed_aging}) with $q = 5 \times 10^{-5}$. Note that all three springs are aged under the strain they undergo when $\epsilon=0.01$ and $x=0.12$. For the weak spring, aging strain is fixed at $|f_0 + A|/10^{-3}$.
%\textcolor{red}{We start with $k = 1$ in the strong springs and $k_w = 9 \times 10^{-3}$ in the weak spring, $\epsilon = 0.01$, and the middle node at $x = 0.12$. We hold the system at fixed strain with $\gamma_w$ at its largest value before aging. Note that the stiffness in the definition of weak force in Eq.~\ref{eq:f_w} is included in $f_0 = k_w x_0$ \textcolor{green}{This description is too detailed and I don't really understand what you are saying.} and $A = k_w \Delta $ where $\Delta$ is the fixed amplitude of change in the strain. We assume $\Delta=0.11$, $\gamma_1 ,\gamma_2 = \delta \ell = 0.007$ for the strong springs, $\gamma_w = 0.131$ for the weak spring, and $ q = 5 \times 10^{-4}$.} %We then update the spring constants according to training protocol in Eq.~\ref{eq:directed_aging}.
After $50$ iterations, we fix the spring constants at the aged values %, $k=0.9993115$ and $k_w=0.008972$, 
and drive the system as in Fig.~\ref{fig:omega_c}a. Because the critical frequency is shifted to a lower value, the middle node can only fall into the global minimum for the lowest frequency shown. It is also possible to increase the critical frequency by aging the system in ways that slow down the dynamics.
%Note the weak spring does not have an assumed rest length; thus we take $\gamma_w(t) = f_0 + A \sin(\omega t)$ as the strain during training. . 

\section{Conclusion}
We have introduced a model that generalizes the notion of hysterons to a dynamical unit that exhibits a variety of distinct behaviors in response to %cyclic 
driving. 
Starting with two springs connected to an overdamped %($i.e.$ massless) 
node, we show that %in the presence of a double-well potential with similar energies,
the system's path is the only factor that determines which well it will choose. Adding a weak spring breaks the symmetry and brings in the energy landscape as another way to determine which minimum is chosen. Thus there is a competition between energy and path for choosing the minimum. %When one of the wells is at a lower energy, it is considered the global minimum. In this case, when the system is driven slowly, the node follows the gradient of energy and can reach the global minimum: a behavior that is energy dependent. However, if the driving is faster than a critical value, the node does not make its way to the global minimum and therefore chooses its equilibrium state based on the path. 
For cyclic driving, %we show that %the frequency of driving can change the dynamics so that 
the system can require multiple cycles to reach its global minimum. 
%This marks the existence of \textit{creep} in the dynamical hysteron model.
% CWL: I probably would've put less of a summary in the conclusion and instead just recapped the results

%The model presented here can be generalized further to explore other aspects of memory formation in elastic materials. 
The model presented is suitable for studying the effect of dynamics in cyclically driven experimental systems where a frequency-dependent behavior is observed. In addition, the sensitivity of the model to directed aging shows its versatility and opens the door to a large variety of training possibilities. 
One natural generalization of this model would be to include interactions between pairs of dynamical hysterons. In the case of ordinary hysterons, the presence of interactions leads to memories that include multiple cycles~\cite{lindeman2021multiple, keim2021multiperiodic}. It would be interesting to see if unexpected behavior emerges by allowing interactions between pairs of dynamical hysterons.

\begin{comment}
Another interesting direction to extend the ideas presented here would be hysterons with more than two wells. For instance, one could design a three-well potential with one massless node and four springs. By setting up the configuration carefully, one can guide the system from the first well (which can be any of three wells) to the second well, then to the third well and back to the first one again, with or without visiting the second well on the way back. This means, to retrieve information about the system's path, one must look at the number of times it visits each well in one cycle.  In this particular case, storing information about the number of times the system visits one particular well (the second one) would be sufficient to retrieve the the path it has taken. 
\end{comment}

\begin{figure}
\centering
\includegraphics[width=6.5cm]{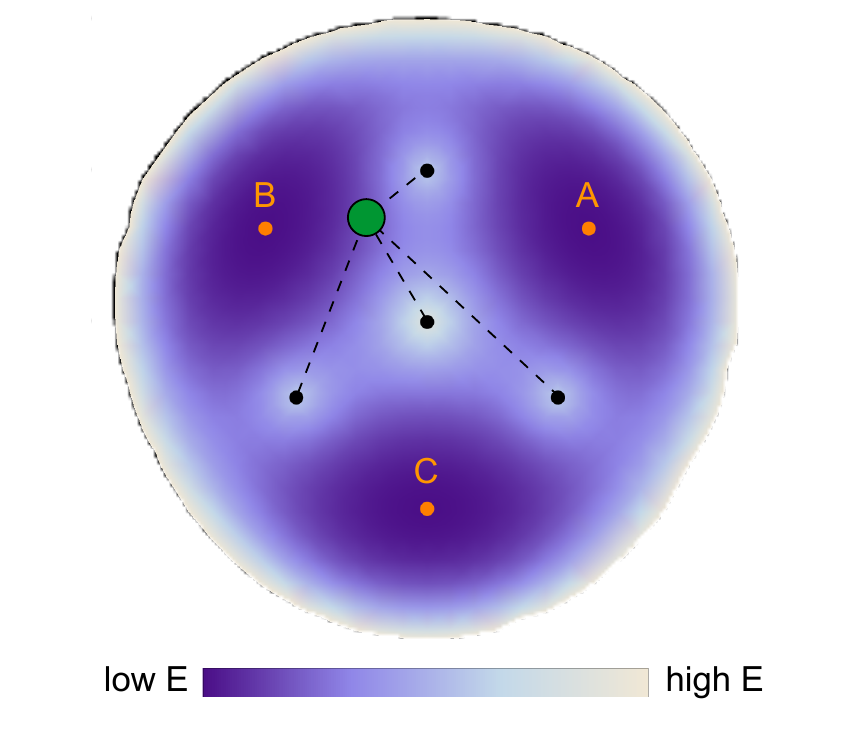}
\caption{A node (green circle) connected by four springs (dashed lines) to the four boundary nodes (black dots). Underlaid is the associated energy landscape for the free node. Orange dots mark the location of the three energy minima. By a suitable reversible manipulation of the boundary nodes, for example as described in the SI, this system behaves as a three-state hysteron that visits the minima sequentially over the course of one cycle: $A \rightarrow B \rightarrow C \rightarrow A$. 
}
\label{three wells}
\end{figure}

Hysterons with more than two wells can also be constructed in a network of springs.  %These offer an additional route to extending the ideas presented here. 
As an example, in Fig.~\ref{three wells} we demonstrate a three-well potential with one %massless 
free node connected to four springs. If the wells are designated $A$, $B$, and $C$, by prescribing the motion of the boundary nodes one can create a reversible boundary motion that causes the middle node to transition from $A \rightarrow B \rightarrow C \rightarrow A$ over the course of a single full cycle (see SI). In this case, the topology of the node's trajectory is nontrivial, a feature that may be important in more complex systems like jammed packings that are sheared repeatedly and where the phase-space trajectory is typically complex (topologically nontrivial) but cyclic.   

We can conceive of using dynamical hysterons in designing mechanical metamaterials. As an example, consider a machine with buckling mechanisms in the form of robotic arms that can switch between two states depending on the frequency. In this case, the dynamical hysterons %would play the role of mechanical logic gates that 
allow changing the structure by controlling the frequency of an external driving force. 

The system presented here is simple enough to be studied analytically, yet the resulting dynamics are rich with complexity. The examples above illustrate that it is easily adaptable to a variety of systems, \textit{i.e.} where interactions between hysterons play a role, or more complex scenarios, as in the case of a three-well system. This dynamical hysteron model can thus provide insight into the effect of new timescales on a wide variety of complex systems.

%\section{Acknowledgements}
We thank Arvind Murugan, Peter Littlewood, and Cheyne Weis for useful conversations. This work was supported by the National Science Foundation (MRSEC program NSF-DMR 2011854) (for model development) and by the US Department of Energy, Office of Science, Basic Energy Sciences, under Grant DE-SC0020972 (for the analysis of aging phenomena in the context of memory formation). C.W.L. was supported by a National Science Foundation Graduate Research Fellowship under Grant DGE-1746045. C.I.I. was supported by the Center for Hierarchical Materials Design (70NANB14H012).

\bibliography{main}

\section{Supplemental information}
\label{sec:SI}
\subsection{Chiral motion in three-well systems}

Using a series of fairly simple motions of the three-well boundary nodes, we can lead the system quasistatically through the cycle $A \rightarrow B \rightarrow C \rightarrow A$. We will use two types of moves: one in which two nodes are pulled away from the center, which causes the two wells opposite those nodes to disappear, and one in which one node is brought toward the center and another is moved azimuthally, which causes one well ($e.g.$ $C$) to remain unaffected and the others to flow ($e.g.$ $A \rightarrow B$). 

Let the top node be node $T$ and the bottom left and right nodes be nodes $L$ and $R$, respectively.

Video 1 shows the following set of steps:
\begin{enumerate}
    \item Nodes $T$ and $R$ move outward. Starting from well $A$, this does nothing
    \item Node $T$ moves toward the center and node $R$ moves azimuthally toward node $L$. This moves us from well $A$ to well $B$
    \item Nodes $L$ and $R$ move outward. This moves us from well $B$ to well $C$
\end{enumerate}
Now these steps are repeated in reverse order. Note that each step had individually reversible boundary conditions, so repeating them in reverse order takes the boundaries back along the same path to the starting point.
\begin{enumerate}
    \item Nodes $L$ and $R$ move outward. We remain in well $C$
    \item Node $T$ moves toward the center and node $R$ moves azimuthally toward node $L$. We remain in well $C$
    \item Nodes $T$ and $R$ move outward. We move from well $C$ back to well $A$
\end{enumerate}
 
 We have therefore moved the boundaries ``out'' and ``back'' along the same path just as you might when shearing a solid out to some maximum strain and back to its original box shape. The result, unlike anything obtainable from a two-well system, is chiral in nature.
\end{document}